\documentclass[acmsmall]{acmart}
\usepackage{subfigure}
\pdfoutput=1
\usepackage[ruled,vlined]{algorithm2e}
\usepackage{amsmath}
\usepackage{float}
\usepackage{graphicx}
\usepackage[utf8]{inputenc} 
\usepackage[T1]{fontenc}    
\usepackage{hyperref}       
\usepackage{url}            
\usepackage{booktabs}       
\usepackage{amsfonts}       
\usepackage{nicefrac}       
\usepackage{microtype}      
\usepackage{lipsum}

\AtBeginDocument{%
  \providecommand\BibTeX{{%
    \normalfont B\kern-0.5em{\scshape i\kern-0.25em b}\kern-0.8em\TeX}}}

\title{Analysis of Agricultural Policy Recommendations using Multi-Agent Systems}

\author{Satyandra Guthula}
\email{guthulasatyandra@gmail.com}
\author{Sunil Easaw Simon}
\email{simon@cse.iitk.ac.in}
\author{Harish C Karnick}
\email{hk@iitk.ac.in}
\affiliation{%
  \institution{Indian Institute of Technology Kanpur}
  \streetaddress{Kalyanpur}
  \city{Kanpur}
  \state{Uttar Pradesh}
  \country{India}
  \postcode{208016}
}

\begin{document}

\begin{abstract}
Despite agriculture being the primary source of livelihood for more than half of India's population, several socio-economic policies are implemented in the Indian agricultural sector without paying enough attention to the possible outcomes of the policies. The negative impact of some policies can be seen in the huge distress suffered by farmers as documented by several studies and reported in the media on a regular basis. In this paper, we model a specific troubled agricultural sub-system in India as a Multi-Agent System and use it to analyse the impact of some policies. Ideally, we should be able to model the entire system, including all the external dependencies from other systems - for example availability of labour or water may depend on other sources of employment, water rights and so on - but for our purpose, we start with a fairly basic model not taking into account such external effects. As per our knowledge there are no available models which considers factors like water levels, availability of information and market simulation in the Indian context. So, we plugged in various entities into the model to make it sufficiently close to observed realities, at least in some selected regions of India. We evaluate some policy options to get an understanding of changes that may happen once such policies are implemented. Then we recommended some policies based on the result of the simulation.
\end{abstract}

\maketitle

\section{Introduction}
\textbf{Motivation:} Socio-economic policy is often formulated and implemented without paying close attention to the possible outcomes of such policy. In the current environment when material conditions change rapidly it is especially important to have appropriate policy responses to avoid adverse socio-economic impact leading to human distress. 

The complexity in strategic management of socio-economic systems lies in the analysis of time evolving structural changes and dynamic aspects of the  system~\cite{lychkina2015agent}. Most policy prescriptions are based on qualitative analysis or simple mathematical models that are often too simple to predict outcomes with confidence. The stochastic nature of the problems and various aspects connected with individual social behavior are very difficult to  capture in such mathematical models\cite{pyka2005agent}. Kremmydas \cite{kremmydas2012agent} describes principles for modelling economic systems using the Agent Based Modelling(ABM) approach and gives its features, advantages and disadvantages. We have chosen the Agent Based modelling approach because of the following two reasons:
\begin{enumerate}
	\item The interactions of the various agents in the system form the basis for the macroscopic equilibrium that the system attains and so following a bottom-up approach, simulating the behaviour and interactions of individual agents is likely to lead to better results compared to a top-down model that may miss out on interactions between agents that affect the dynamics and the final state of the system.
	\item The system can display ``emergent properties''. Bar Yam (1997)\cite{bar1997dynamics} distinguishes two types of emergence, local and global. A phenomenon is emergent if it requires new categories to describe that are not required to describe the behavior of the underlying components. Local emergence is met in low complexity systems where the emergent properties are maintained by some parts of the system as defined in~\cite{byrd2000trust}. On the other hand, global emergence is present only in the system as a whole. For example, in our simulations when farmers were not allowed to interact with each other, each tried to optimize by choosing actions based solely on the choices available to them. But when interaction of Type1 and Type3 farmers was allowed (they are defined in section 3), a Type3 farmer would lend water to a Type1 farmer and also dictate the crop to be produced\cite{watermarket}. This way, the farmers would mass produce a commodity, making more room for profit and improved payment security. 

\end{enumerate}

In the past few years, several attempts have been made to model different socio-economic systems using agent based modelling~\cite{happe2006agent}, like climate change adaptation~\cite{troost2015climate}, pension system in Russia, urban planning and so on\cite{motieyan2018agent}. Multi-Agent Simulation was also used in the analysis of food supply chains where emphasis was given to environmental impact and agent decision processes~\cite{krejci2012modeling}, or analysing the impact of agricultural policies on combination of geophysical and social models specific to a geographic location~\cite{lobianco2010regional}. But the decision process and the interaction among agricultural sector agents varied greatly depending on the traditions and practises prevalent in the region of interest. In India, this kind of modelling is not yet used. Due to different conditions environmental, social and economic the strategies available to farmers vary significantly. The viability of farming depends on the strategies and payoffs for actions available to the agents, which can be verified from the results presented in this paper. We wanted to simulate various scenarios where the concerned parties interacted with each other and how these interactions were affected due to the implementation of certain policies. We also wanted to understand how accessibility to information affected the business of small scale farmers and wanted to understand the efficacy of policies implemented by the Government of India in the agricultural sector. We have chosen a relatively autonomous agricultural sub-system in the sugarcane supply chain as our target system. In our simulations we have analysed results by varying import prices, export prices, FRP (Fair and Remunerative Price) of sugarcane, and the ethanol requirement for the Ethanol Blending Program for auto fuel. We have also done simulations where we vary the information accessible to various sections of farmers and consider various crop alternatives that farmers can use based on the region. 

\section{Overview of the Indian Sugarcane industry}
In 2016-17 the production of sugar was approximately 20.3 million tons when domestic consumption was approximately 24.5 million tons - a shortfall of about 4.2 million tons. This lead to increase in sugar prices and sugar was imported to control prices. The cane farmers responded by increasing the plantation of sugarcane and in the 2017-2018 and 2018-2019 the total sugar produced was 32.5 million tons and 35.9 million metric tons, which led to a crash in sugar prices in both the years\cite{statsindsugar}. International sugar prices were also depressed, ranging between Rs 22-Rs 26 per kg in 2017-2019 so even the export option was not available\cite{commodityprofsug}. In 2019-2020 the government mandated FRP for sugarcane was Rs.275/- per Quintal (100Kg) and in Uttar Pradesh it was Rs.325 per Quintal. FRP is the government mandated price at which the sugar mills buy sugarcane from farmers. If the mills do not have enough funds to buy sugarcane from the farmers, then cane is borrowed and the due amount is calculated with the current FRP, which is payable later when the funds are available. MSP (Minimum Support Price) is the price at which the Government of India buys crops or farm produce directly from farmers. MSP is applicable only to certain types of crops. Due to high FRP, the sugar mills are not able to pay the farmers and this has led to a piling up of sugarcane payment dues to farmers to the tune of Rs.220 billion \cite{statsindsugar}. This is an economically disastrous situation for sugar mills and and has meant wide spread economic distress for sugarcane farmers. The most recent sugarcane farmer protest happened on 16\textsuperscript{th} December 2019 urging the State Government of Haryana to increase the prices of sugarcane\cite{haryanaprices19}. Under the Ethanol Blended Petrol program, the blend targets have been increased from 5 \% to 10\% in 2019\cite{dfpd}. We analysed the outcomes of the policy changes introduced by the Central Government of India and included them in the results.

\section{Our Approach}
We have modelled the sugarcane industry as a Multi-Agent System with the following agents. The behaviour of the agents and their attributes are based on what we have been able to find out about the sugar industry from existing published or online sources.
\begin{enumerate}
\item \textbf{Type1 Farmer} Farmer with low savings (1 million rupees) and 1.5 ha (hectare) land for agricultural use. The water resources are limited and in some cases not sufficient for farming certain varieties of crops. Around 70\% of the farmers belong to this category. The information present with these farmers is very limited. Information is modelled as a Gaussian with mean as the actual value being queried and standard deviation depending on the type of farmer.\cite{criteriafarmers} 
\item \textbf{Type2 Farmer} Farmer with average savings (3 million rupees) and 3 ha land for agricultural use and sufficient water sources for farming any kind of crop. The information present with these farmers is quite reliable with unit standard deviation\cite{criteriafarmers}
\item \textbf{Type3 Farmer} Farmer with average savings (5 million rupees) and 4.5 ha land for agricultural use, abundant water sources which can be lent to Type1 Farmers in exchange for a specific variety of a crop produce\cite{watermarket}, good sources of information regarding sales, produce, import/export. This type of farmer has some income from Type1 farmers and can influence mills and cold storage owners due to their ability to dictate the crop to be produced by type1 farmers, this agent is crucial for coordination among type1 farmers. They are assumed to have perfect information. \cite{criteriafarmers}
\item \textbf{Water agent} Agent that sells water for a price, unlike the Type3 farmer where the crop is dictated and collected as payment. They are controlled by government, so policies can be imposed.\cite{watermarket}
\item \textbf{Loan agent} Agent that provides working capital based on collateral or a credit score. The collateral based loans have 8\% rate of interest, but credit based loans have 12\% rate of interest. In general, the collateral based loan interests are paid with higher preference because after 4 defaults the collateral is confiscated and the difference is paid back to the customer.
\item \textbf{Mill agent} Agent that accepts the sugarcane produce, and generates sugar or alcohol. This agent accepts the produce, but pays when there are enough funds to pay the farmers. Till then the due is recorded. The mill agent goes to the farmers only if a threshold produce is met.
\item \textbf{Storage agent} Agent that accepts the produce to be stored and retrieved when necessary by the farmers. Type3 Farmers have influence on this agent, so they are given higher preference followed by Type2 farmers and Type1 Farmers. The storage cost and spoilage of commodities is dependant on the type of crop.\cite{coldstorage}
\item \textbf{Market agent} Agent that sets the price of the commodities based on demand, supply and commodity. All the sale happens via this agent. The price is altered based on the trend of the sales and commodity stock available, so we check how much the sale in each cycle varied from the usual sale. Using this sale difference and with higher priority given to the recent cycles, we proportionately alter the price based on the past 4 cycles. For example if the sale differences in the past 4 cycles are -5\%, +5\%, +10\%, +2\%, we do $0.4*2\% + 0.3*10\% + 0.2*5\% -0.1*5\%$. So, we increase the current price from the previous price by 4.3\%. 
\item \textbf{Consumer agent} This is the agent that consumes the commodity and creates demand. The need for each commodity is within a specified limit irrespective of the price of the commodity, for example sugar consumption from 2015-2019 has been in the range of 24.5 million tons to 26 million tons per year\cite{consumption}. Even with very low sugar prices and high availability of sugar, the demand cannot rise dramatically because human consumption has an upper bound. In 2018 19.5 kg of sugar was recorded as the per capita consumption. So, the total sale of sugar within past 4 cycles is used to calculate the demand for the current cycle. We have chosen a cycle as 3 months in this example, so in each cycle the usual demand should be 4.9 kg. So, if the current price is 10\% lower than the previous cycle, and the sugar purchased in the past 4 cycles is 20\% lower than the usual demand for the 4 cycles (19.5 kg), then the current demand would be $1.1*1.2*4.9kg$. This is because due to 10\% decrease in price, we expect 10\% increase in demand, and because there was 20\% less sugar purchased previous year, we expect shortage in the stockpiled sugar and the demand to go up by 20\%. The Indian market is highly price sensitive.
\end{enumerate}
Due to abundance of labour in India, the labour charges are considered fixed, and increase solely by inflation. Type1 farmers do not have sufficient information and hence their choices rarely give the expected outcomes. But due to the influence of type3 farmers, the type1 farmers planted sugarcane and chose to keep planting sugarcane even after the farmers started getting their water from a water agent and not a type3 farmer. Hence the type3 farmers are crucial, using their information they choose to maximise their profit by forcing the type1 farmers to produce crops that would yield higher income than other crops, which the type1 farmers rely on in future time cycles. The following figures \ref{fig:oa} to \ref{fig:millai} show the interactions among agents. 
\begin{figure}[H]
 \begin{center}
  \includegraphics[scale=0.5]{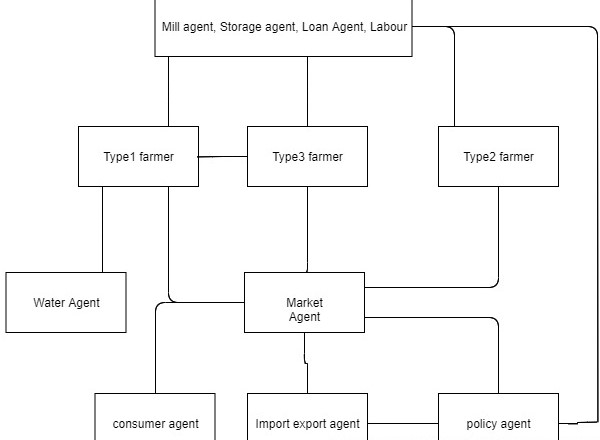}
  \vspace{-0.5em}
  \caption{\small{A basic template view of interactions among agents. The interactions of each agent are shown separately later.}}
   \label{fig:oa}
 \end{center}
 \vspace{-1.5em}
\end{figure}

\begin{figure}[H]
 \begin{center}
  \includegraphics[scale=0.5]{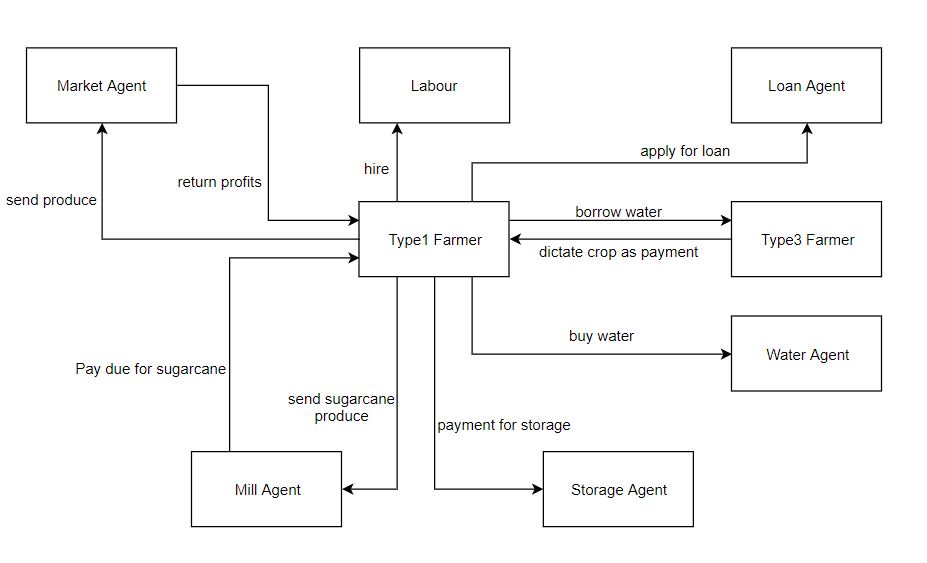}
  \vspace{-0.5em}
  \caption{\small{This figure shows how the Type1 farmer interacts with other agents}}
   \label{fig:t1i}
 \end{center}
 \vspace{-1.5em}
\end{figure}

\begin{figure}[H]
 \begin{center}
  \includegraphics[scale=0.5]{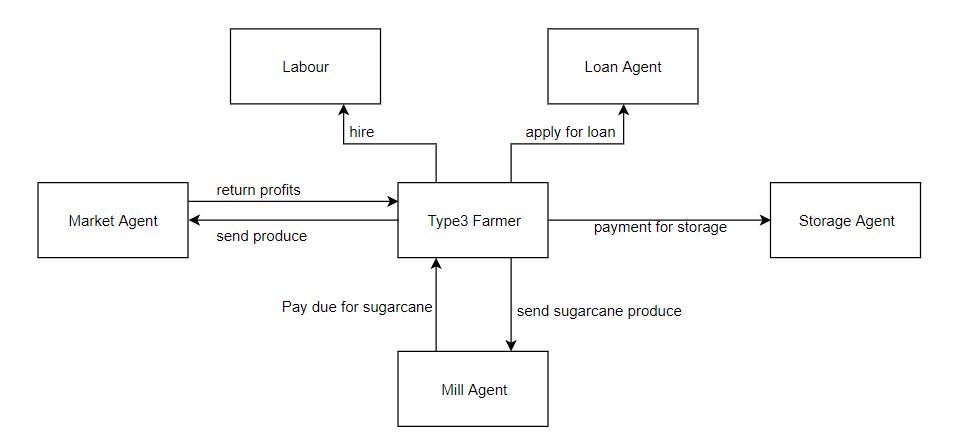}
  \vspace{-0.5em}
  \caption{\small{This figure shows how the Type2 farmer interacts with other agents}}
   \label{fig:t2i}
 \end{center}
 \vspace{-1.5em}
\end{figure}

\begin{figure}[H]
 \begin{center}
  \includegraphics[scale=0.5]{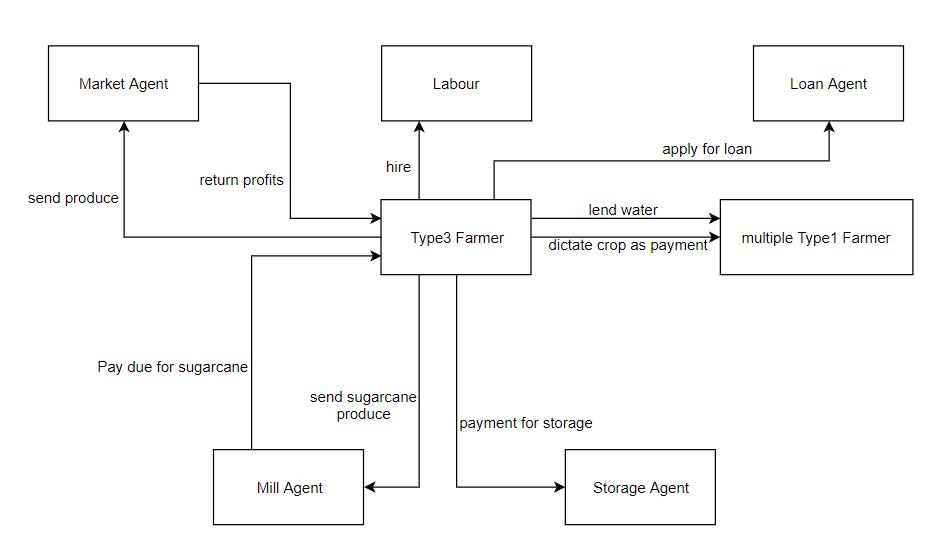}
  \vspace{-0.5em}
  \caption{\small{This figure shows how the Type3 farmer interacts with other agents}}
   \label{fig:t3i}
 \end{center}
 \vspace{-1.5em}
\end{figure}

\begin{figure}[H]
 \begin{center}
  \includegraphics[scale=0.4]{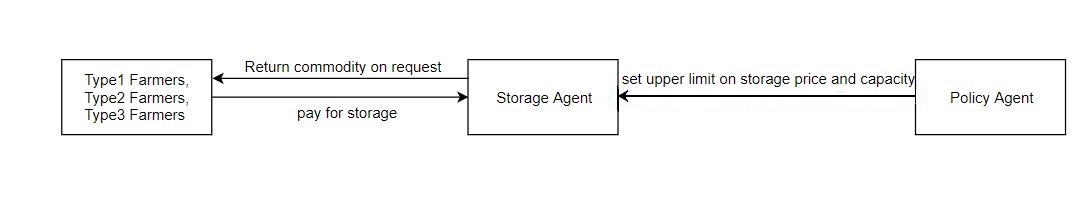}
  \vspace{-0.5em}
  \caption{\small{This figure shows how the Storage agent interacts with other agents}}
   \label{fig:sai}
 \end{center}
 \vspace{-1.5em}
\end{figure}

\begin{figure}[H]
 \begin{center}
  \includegraphics[scale=0.4]{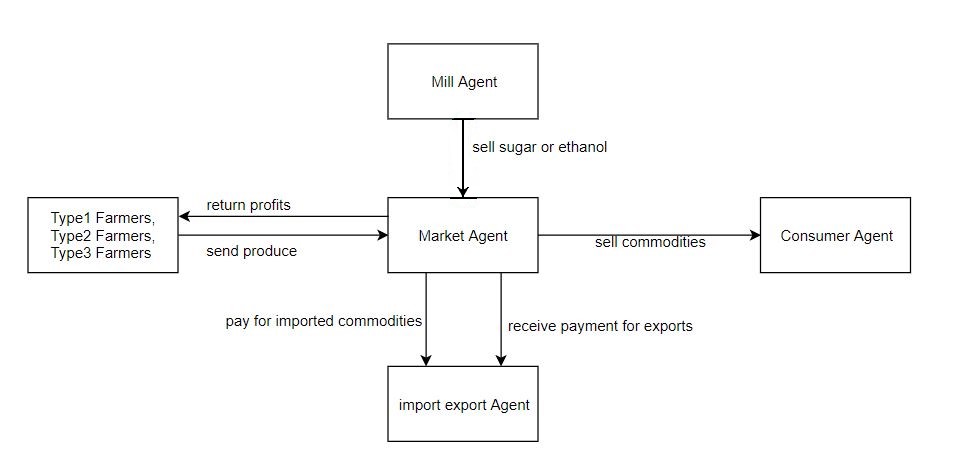}
  \vspace{-0.5em}
  \caption{\small{This figure shows how the market agent interacts with other agents}}
   \label{fig:mai}
 \end{center}
 \vspace{-1.5em}
\end{figure}

\begin{figure}[H]
 \begin{center}
  \includegraphics[scale=0.5]{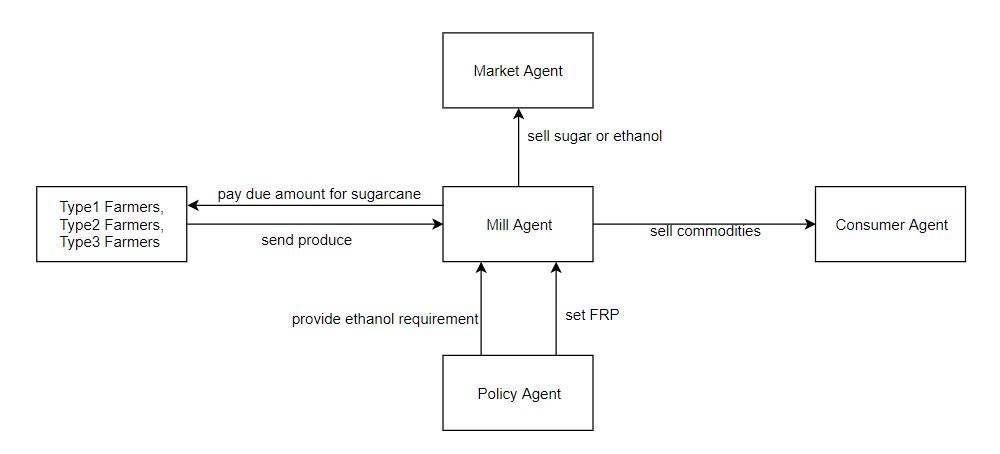}
  \vspace{-0.5em}
  \caption{\small{This figure shows how the mill agent interacts with other agents}}
   \label{fig:millai}
 \end{center}
 \vspace{-1.5em}
\end{figure}
\section{Results}
We will see how some of the most common policies implemented  by the government affect the system. We also show some alternate policies that give some good results in the simulations.\newline

First we define what is meant by good results. In the simulations we have set 70\% of the farmers to be of Type1, 20\%  as Type2 and remaining as Type3~\cite{percentagefarmers}. We have simulated using 10,000 farmer agents. Now, if the total savings goes below a certain number, a family cannot be sustained, so in that case the farmer ``moves out of farming" and finds an alternate source of income. So, in the simulations if a farmer's savings goes below the cumulative per-person-charge for a family for 4 time steps then the farmer stops farming and moves to a different line of work. Ideally, the number of farmers moving out should be 0 assuming no other external influences, which means farming is sustainable. But, that is not the case, the simulations show that without proper help the number of farmers moving out can go as high as 80-90\%. So, in the simulations we use number of farmers moving out as the metric for testing the efficacy of a policy.

We will see how the simulations run for the current real world values for the parameters like FRP, demand and export prices.

\begin{figure}[H]
	\begin{center}
	\subfigure[]{
		\includegraphics[scale=0.26]{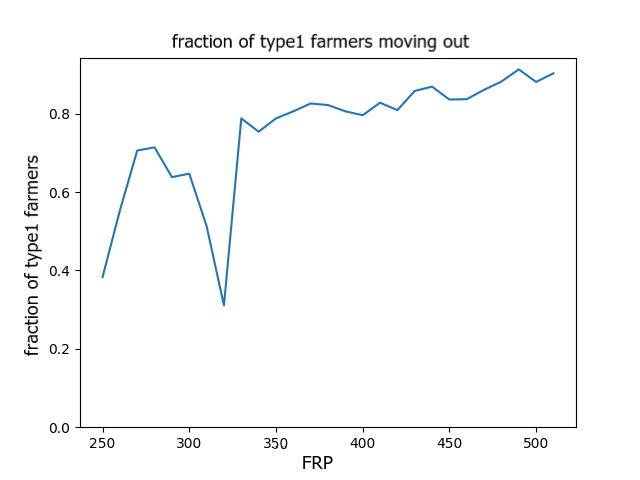}
		\label{fig:movefrp}
		}\hspace{0.2cm}
	\subfigure[]{
		\includegraphics[scale=0.26]{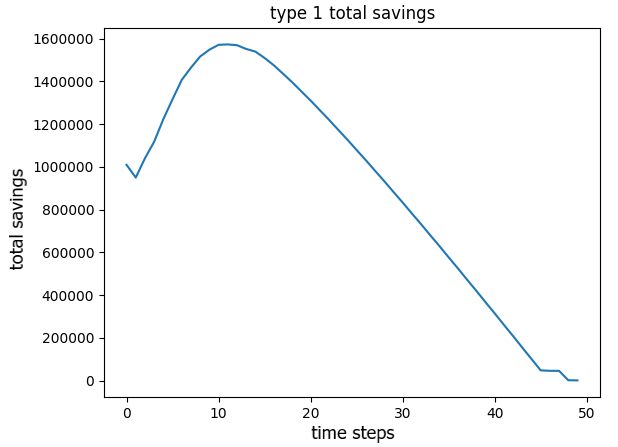}
		\label{fig:lowfrp1}
		}\hspace{0.2cm}
		\subfigure[]{
		\includegraphics[scale=0.26]{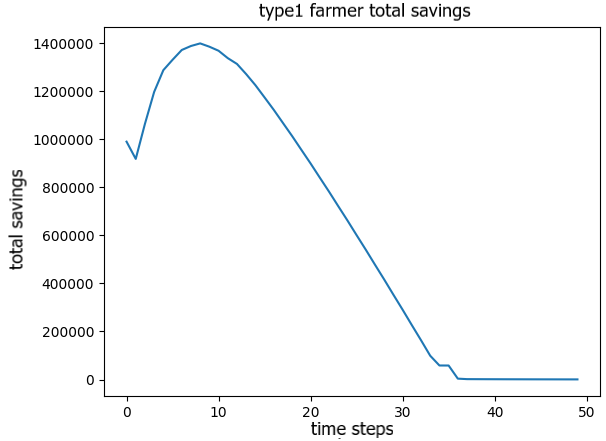}
		\label{fig:highfrp}
		}
	\end{center}
	\caption{\small{(a) Graph of the number of farmers moving out vs FRP for sugarcane in Rs/Quintal (b) Graph of the savings of Type1 farmers vs time steps when FRP is 220 per Quintal (c) Graph of the savings of Type1 farmers vs time steps when FRP is 480 per Quintal}}
\end{figure}
As we can see from Figure \ref{fig:movefrp}, changing the FRP for sugarcane does not help the farmer. Rather higher values of FRP increases the number of farmers moving out of sugarcane farming. This is because, FRP is the initial income expectation. Due to it being high, most of the sugarcane farmers plant sugarcane. Now, if the FRP is low, after the first few time steps, the farmers deem the income unsustainable and stop planting it all together. But with high FRP, the number of time steps before the income expectation falls from FRP to a non profitable income is higher. Even the mill owners keep track of the sugar already existing and hence they do not buy if they have excess sugar. So, when the farmers keep producing sugarcane and the mill owners buy very limited amounts of sugarcane it leads to the farmers dumping sugarcane that makes the situation worse. We can see this effect from Figures \ref{fig:lowfrp1} and \ref{fig:highfrp}. In 2014, sugar price fell to Rs.24 per kg, in the crushing season of 2015, several mills in Uttar Pradesh refused to buy sugarcane from farmers. After the government made it mandatory for mill owners to buy sugarcane from farmers, the arrears of 2016-2017 reached Rs.25.35 billion.

\begin{figure}[H]
	\begin{center}
	\subfigure[]{
		\includegraphics[scale=0.3]{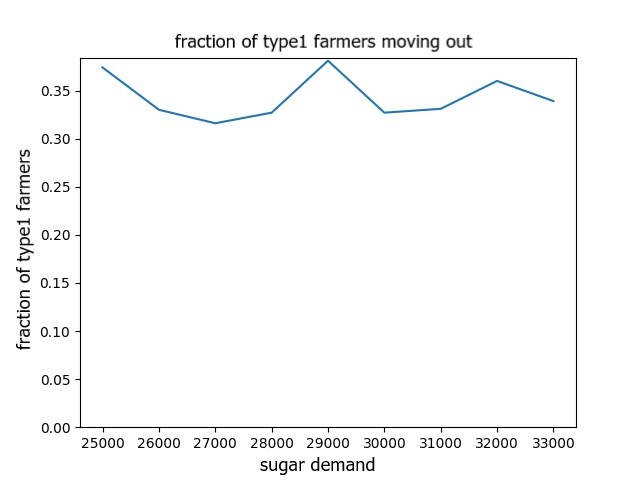}
		\label{fig:movedemand}
		}\hspace{0.2cm}
	\subfigure[]{
	    \includegraphics[scale=0.3]{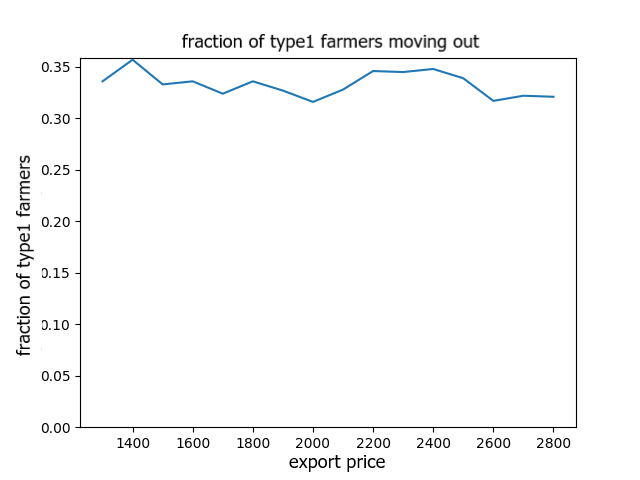}
		\label{fig:moveexport}
	}
	\end{center}
	\caption{\small{(a) Graph of the number of farmers moving out vs demand of sugar in Quintals (b) Graph of the number of farmers moving out vs export price of sugar in Rs/Quintal}}
\end{figure}

From figure \ref{fig:movedemand} we can see changing the demand is not helpful either, unless the demand is very high. As the price is very low rationally the market simply exports the sugar. But since export prices are low it does not boost incomes which does not help in reducing the stress. 

\begin{figure}[H]
	\begin{center}
	\subfigure[]{
		\includegraphics[scale=0.3]{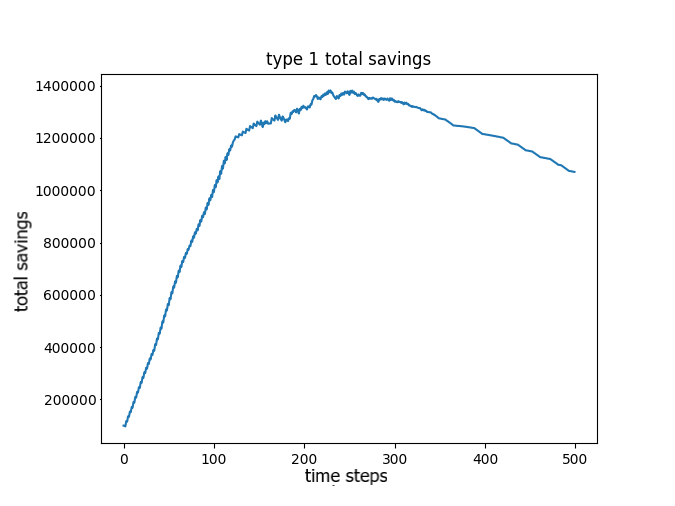}
		\label{fig:t1storage}
		}\hspace{0.2cm}
	\subfigure[]{
	    \includegraphics[scale=0.3]{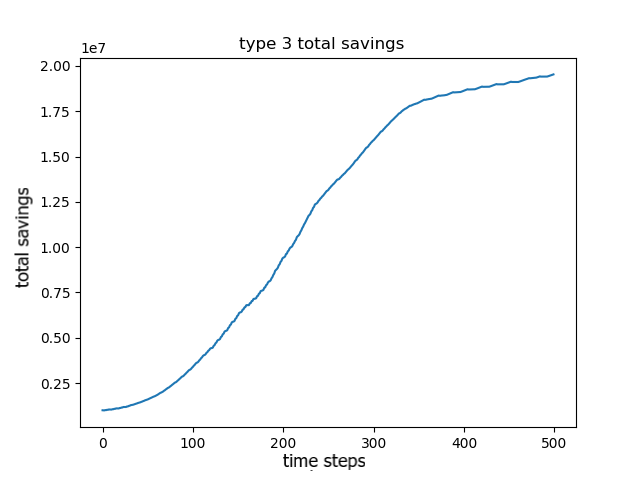}
		\label{fig:t3storage}
		}
	\end{center}
	\caption{\small{(a) Graph of the savings of Type1 farmers vs Number of time cycles(number of 4 month cycles) (b) Graph of the savings of Type3 farmers vs Number of time cycles(number of 4 month cycles}}
\end{figure}

From figure \ref{fig:moveexport}, we can see that without much increase in export price not much difference can be seen in the number of farmers exiting from farming. The reason is the same as  before. If the export price is significantly higher then most of the produce is exported leading to increased prices which depresses demand and the number of farmers moving out becomes higher. \\
The increase in storage facilities do not help increase the income of Type1 farmers significantly in general. But there was a significant increase in the case of Type3 farmers. This is not specific for sugarcane farming. Figures \ref{fig:t1storage} and \ref{fig:t3storage} show how the Type1 farmers' savings and Type3 farmers' savings changed over the course of time when the storage capacities increased.

Below are some simulations that reduced the number of farmers moving out of sugarcane farming: 
\begin{enumerate}
	\item By increasing the requirement of ethanol we can see some substantial improvement in the income of the farmers. Figure \ref{fig:moveethrq}  and \ref{fig:moveethrq10} show the number of people moving out of sugarcane farming versus the ethanol requirement for blending. At present blending is only 3.5\%. The simulation shows that if blending of 10\% is made mandatory, then approximately 50\% of the sugarcane farmers will remain in farming. By changing the blend percentage to 20\%, 70\% of sugarcane farmers will remain in farming. We also tried to see whether changing the price of ethanol helps sugarcane farmers. Figure \ref{fig:moveethp} shows the results.
	\begin{figure}[H]
		\begin{center}
		\subfigure[]{
			\includegraphics[scale=0.26]{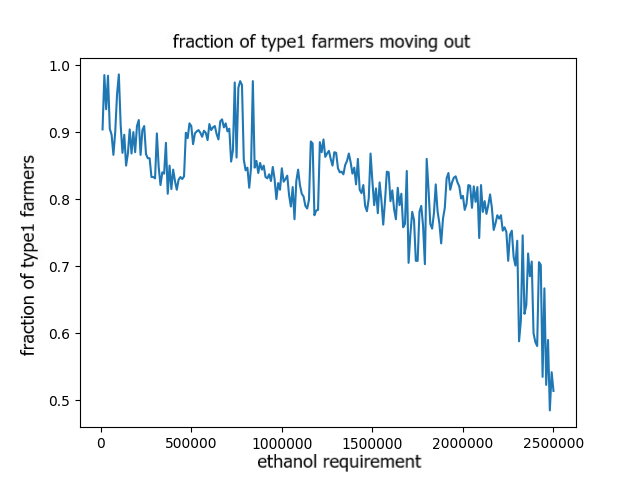}
			\label{fig:moveethrq}
			}
		\subfigure[]{
		    \includegraphics[scale=0.26]{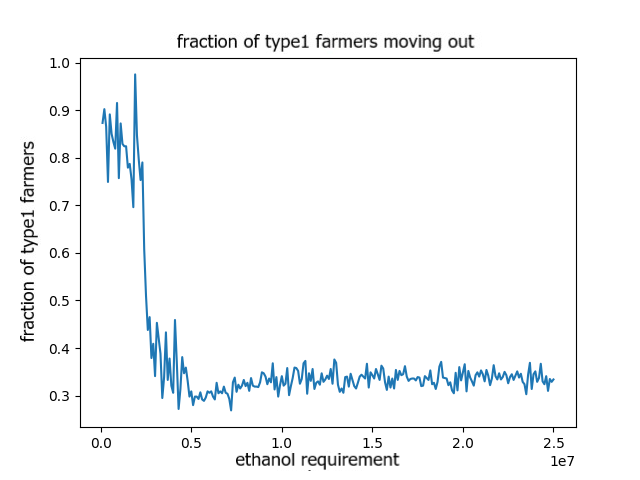}
			\label{fig:moveethrq10}
		}
		\subfigure[]{
		    \includegraphics[scale=0.26]{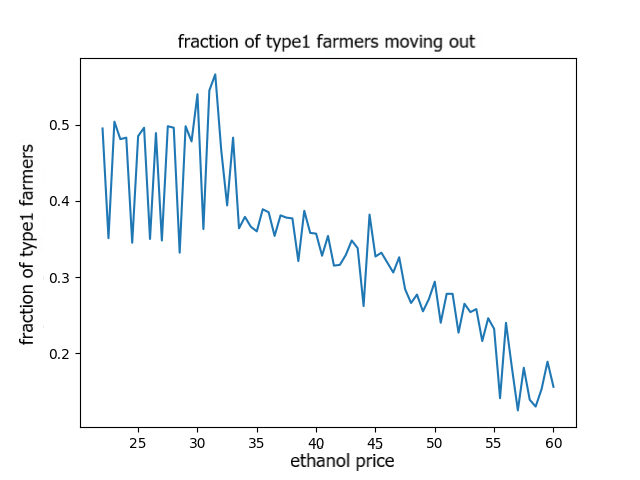}
			\label{fig:moveethp}
		}
		\caption{\small{(a) Graph of the number of farmers moving out vs requirement of ethanol in kilo-litres, (b) Graph of the number of farmers moving out vs requirement of ethanol in kilo-litres with x-axis scale 10 times as (a), (c) Graph of the number of farmers moving out vs price of ethanol in Rs/L}}
		\end{center}
	\end{figure}
	
	\item Increasing ethanol requirement or increase in ethanol price also leads to stabilization of price of sugar. Due to ethanol business being sustainable sugar production can be controlled, as opposed to converting all the sugarcane to sugar due to the wastage of sugarcane involved. Figure \ref{fig:priceinstabprice} captures this effect. In all the graphs, `Price of Type 0' refers to sugar price.
	\begin{figure}[H]
		\begin{center}
		\subfigure[]{
			\includegraphics[scale=0.3]{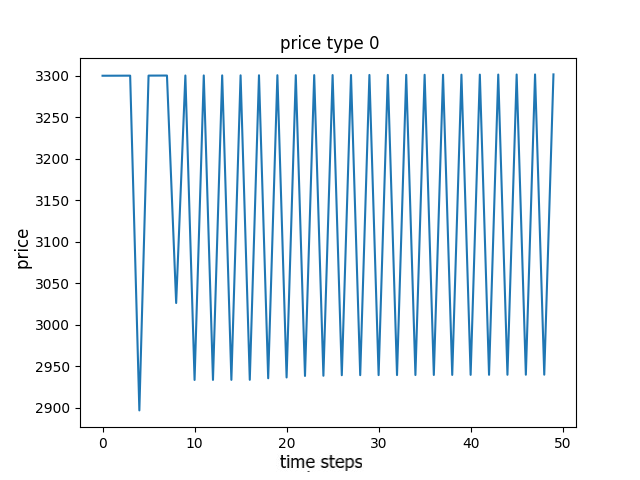}
			\label{fig:priceinstabprice}
			}
		\subfigure[]{
		\includegraphics[scale=0.3]{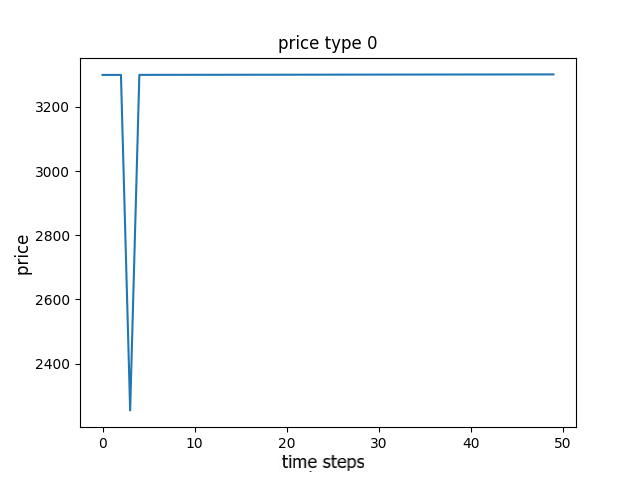}
			\label{fig:pricestabprice}
		}
		\end{center}
		\caption{\small{(a) Graph of price of sugar in Rs/Quintal vs Time cycles when ethanol price is kept as 22 Rs/L. Though the fluctuation may seem huge but the fluctuation is actually small (4 Rs/kg) (b) Graph of price of sugar in Rs/Quintal vs Time cycles when ethanol price is kept as 58 Rs/L.}}
	\end{figure}
	\item From figure \ref{fig:pricestabprice}, it can be seen that the price does not change a lot. The initial price fluctuation is due to excess sugar produced due to lack of initial demand information for sugar. Because sugar can be stored easily, the amount of sugar in the market can be monitored and the price of sugar kept in check. The increase in ethanol requirement promotes the production of ethanol and hence does not let the store houses be filled with unmanageable amount of sugar.
	\begin{figure}[H]
		\begin{center}
		\subfigure[]{
			\includegraphics[scale=0.3]{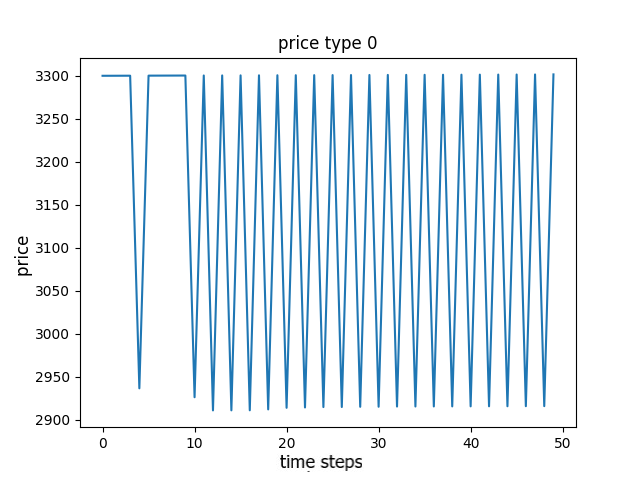}
			\label{fig:priceinstabreq}
			}
		\subfigure[]{
		\includegraphics[scale=0.3]{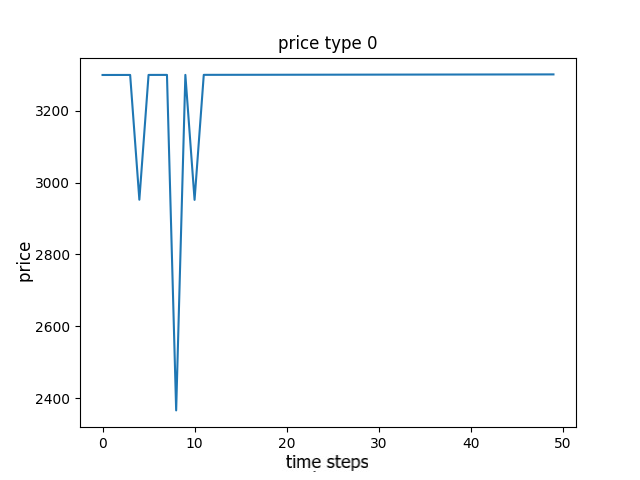}
			\label{fig:pricestabreq}
			}
		\end{center}
		\caption{\small{(a) Graph of price of sugar in Rs/Quintal vs Time cycles when ethanol requirement at is kept at 40,00,00,000 Litres. (b) Graph of price of sugar in Rs/Quintal vs Time cycles when ethanol requirement is kept at 2,20,00,00,000L. From the graph, it can be seen that the price doesn't change a lot here as well}}
	\end{figure}
	\item From the figures \ref{fig:priceinstabreq} and \ref{fig:pricestabreq} we can see that the increase in ethanol requirement can also provide a stabilizing effect on the price of sugarcane. Even at low prices of ethanol, if the demand for ethanol is significant the total profit made by the mill owners is substantial and makes the ethanol business viable. So, similar to the case where the ethanol prices were changed, the sugar production can be manipulated to make the overall sugar and ethanol business profitable.
		\item We have also tried to see what happens when farmers are given an alternative crop to sugarcane. With the introduction of one alternative whose price is one third the price of sugar and export price slightly higher than the average price of the new crop (10\% more) the situation improves greatly. The main advantage here is that the ratio of people moving out of  agriculture becomes approximately 5\% which is much better than the best case scenario of ethanol price increase or  ethanol requirement increase (which is approximately 20\%).
	\item The addition of the new crop also has a stabilizing effect on the prices of both  crops. 

\end{enumerate}
With the introduction of a second crop and alternating the crops the farmer can get better income because by the time the second crop gives a yield the price of sugar goes up and vice-versa. \newline \newline
So, the following are some policy suggestions:
\begin{itemize}
	\item Increase the ethanol requirement by increasing blending percentage of ethanol with petrol.
	\item Increase the ethanol price to a value above 55 or 60 Rs/L. This will be a small price to pay if we can treat petrol just as another commodity and apply GST.
	\item Come up with alternative crops for sugarcane farmers that have a produce price at least 5 times the FRP, export price at least 10\% more than the average price offered for the new crop and the yield of at least 5 tonne per acre, which is 10\% of the sugarcane yield per acre. For example onions or potatoes.
\end{itemize}

We have also simulated the policies implemented by Central Government of India regarding sugarcane.\newline
We have simulated 10\% ethanol blended with petrol policy by varying the FRP from Rs 270 per Quintal to Rs 500 per Quintal. The ethanol prices used were Rs 51 per litre and Rs 60 per litre, because these were the prices proposed by the Central Government depending on how the ethanol was made in the mills. 
\begin{figure}[H]
		\begin{center}
		\subfigure[]{
			\includegraphics[scale=0.3]{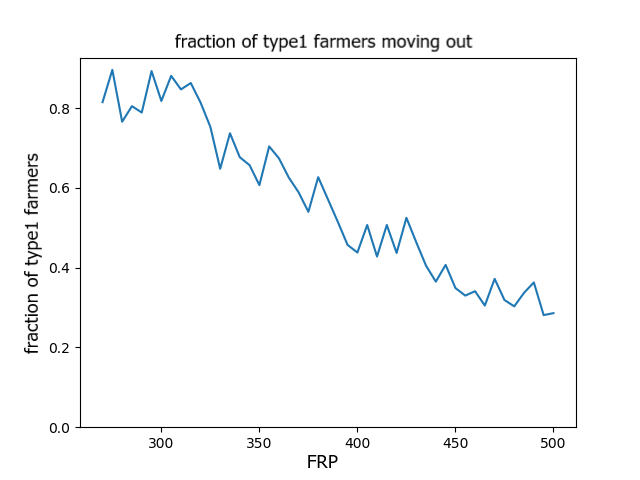}
			\label{fig:changingfrpeth51}
			}
		\subfigure[]{
		\includegraphics[scale=0.3]{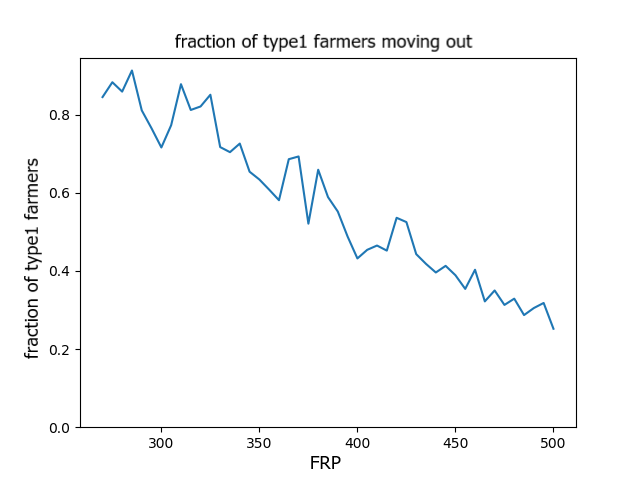}
			\label{fig:changingfrpeth60}
			}
		\end{center}
		\caption{\small{(a) Graph of Farmers leaving sugarcane agriculture vs FRP when ethanol price is Rs 51/litre. (b) Graph of Farmers leaving sugarcane agriculture vs FRP when ethanol price is Rs 60/litre.}}
\end{figure}

\begin{figure}[H]
		\begin{center}
		\subfigure[]{
			\includegraphics[scale=0.3]{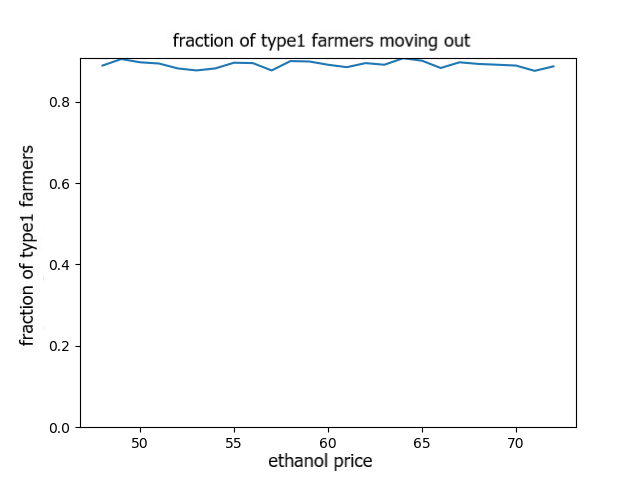}
			\label{fig:ethchange}
			}
		\subfigure[]{
		\includegraphics[scale=0.3]{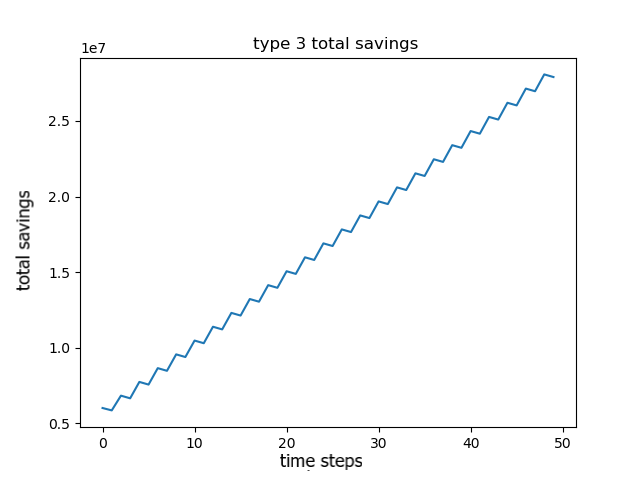}
			\label{fig:t3lowfrp}
			}
		\end{center}
		\caption{\small{(a) Graph of Farmers leaving the sugarcane agriculture vs ethanol price when FRP is Rs 275/Quintal. (b) Graph of type 3 farmer's savings vs time cycles when FRP is Rs 275/Quintal and ethanol price is Rs 51/litre.}}
\end{figure}

From figures \ref{fig:changingfrpeth51} and \ref{fig:changingfrpeth60}, we see that unless the FRP increases the value paid by the mills to the farmers it does not sustain the farmers. From figure \ref{fig:ethchange} we can see even after changing  ethanol prices, if the FRP remains the same, it is still not profitable for type1 farmers. But for type3 farmers who have large land holdings, sugarcane farming remains profitable, which can be seen from figure \ref{fig:t3lowfrp}.

\section{Conclusion}
Multi agent systems can be used as a reliable tool to simulate the outcomes of different policies. While the current model is fairly simple it is much easier to build more detailed models with more complex decision processes for  agents and richer interactions among agents. More work needs to be done on using Multi-Agent Systems coupled with intelligent learning agents in the socio-economic sector. The behaviour of the agents can be changed and the number of agents can be scaled depending on requirements and resources. Availability of good data and the ability to identify and model all the agents and factors in the real system is the main limitation for this kind of modelling. Assuming the model is valid the simulations show that current policies are not sufficient for improving the livelihood of farmers. The suggested alternatives should be considered and empirically tested.

\bibliographystyle{ACM-Reference-Format}  
\bibliography{reference-list}  


\end{document}


\maketitle

\section{Supplementary}

Here we describe the approach we have used to model the sugar-sugarcane system as a multi-agent system, the assumptions that were made and the actions of each agent.
\subsection{Description of Agents and Objects used}
Here we describe the actions each agent can take, the parameters or attributes of each object an agent will interact with and the decision process of each agent.\\\\
Now, for simplicity, we have not considered agricultural seasons because taking optimum decision of which crop to plant or farm  would depend on the current season, when you can plant the same or a different crop again, when will the crop give produce and so on. So, we have taken crops that can be planted at the same season. So, for this approach to work the time step for all the crop or land related activities should be something that will not involve seasons. So, we have taken the time step as the time of the shortest harvest-cycle among the harvest-cycles of the crops. Harvest cycles will be explained later in this section.\\\\
First let us look at the parameters of some objects the agents will deal with.
\subsubsection{Objects}
\begin{enumerate} 
	\item \textbf{Crop objects}
A crop object has the following attributes:
\begin{itemize}
	\item End-cycle: tells us the number of time-steps by which the planted crop becomes unusable, or not profitable. This will be different for each crop. For instance, if the time step used in the simulation is 4 months then if a crop can produce fruits in profitable quantity and quality for 10 years, then we say that the value for the end-cycle attribute is 30. This is purely a property of the crop.
	
	\item Harvest-Cycle: Number of time steps necessary before yielding fresh produce after the previous harvest or produce. It does not take into account the time for the plant to grow into a tree before it yields the produce, it only measures the time taken after the previous harvest. Though it is simple to do that by adding another variable we have not done it because the crops we were testing were the crops where the harvest-cycle and end-cycle were the same. For instance. If the time step is 1 month, then if we have a crop that gives produce once every year, then we say the harvest-cycle is 12. In the case of sugar cane, both the harvest-cycle and end-cycle are the same.

	\item Fertilizer and pesticide: Cost of fertilizers and pesticides for a crop per unit land.
	\item  labor-requirement: The optimum labor requirement for a crop per unit of land.
	\item Water-requirement: The optimum water requirement for a crop per unit land.
	\item labor-flexibility: The difference in labor can change the quality of the crop. For now, we have assumed that the crop's quality is linearly dependent on the ratio of the labor present and labor-requirement for each crop. The attribute labor-flexibility gives us the weight of the above ratio.
	\item Water-flexibility, which tells us how the difference in water given to the crops can change the quality of the crop. The water-flexibility is similar to labor-flexibility and has the same assumptions.
	\item Prone-to-pest: Factor that tells us before how many time steps of not using pesticides does the plant die.
	\item Produce: The amount of produce that can be estimated per unit land occupied by a crop under ideal conditions, i.e when the crop has all requirements satisfied.
	\item Initial cost: The one-time cost of each crop, such as seed cost, equipment cost.
	\item Minimum produce (optional): Specifies the minimum amount of product to be made before which the intermediate agents such as mill agents can consider taking the produce from the farmer.
\end{itemize}
\item \textbf{labor objects}
labor is not treated as an agent because of the abundance of labor in India - though this is gradually changing in parts of the country. But it is still substantially true for agriculture.
\begin{itemize}
	\item Wages: Specify the wages for each crop after the crop has been planted per unit land per unit time. This also increases as the time progresses by a factor of 0.0001(inflation).
\end{itemize}
\item \textbf{Storage object}
\begin{itemize}
	\item Capacity: The storage capacity for each crop.
	\item Fee-multiplier: The fee-multiplier is an attribute specific for storing each crop. So, based on the storage space remaining and the amount of storage requested by all the agents, the fee is modified. This increases every time step by 0.0001(inflation)
	\item Loss-rate: The amount of produce stored at the current time step that will be spoilt by the next time step.
	\item Expiration date: The time by which the stock will be unusable.
\end{itemize}
\end{enumerate}
\subsubsection{Agents}
\begin{enumerate}
	\item \textbf{Type1 Farmer agent}
		\begin{itemize}
			\item Each Type1 farmer has a family size that is drawn from a uniform distribution between 4 and 6.\cite{criteriafarmers} 
			\item Each agent has Total Savings drawn from a gaussian with mean as 5 lakh rupees and standard deviation as 10000 at the initialization phase of the agent\cite{criteriafarmers} 
			\item Each agent has a parameter called per-person-charge which is the minimum amount of money needed to survive. This is a constant value of 5000 rupees per month, which increases as the simulation progresses with a factor of 0.0001 (inflation)
			\item Each agent has a safety buffer as 10\% of initial savings. It keeps increasing by a factor of 0.0001 with each time step
			\item Each agent has land as an attribute which is drawn from a gaussian with mean as 1.5 ha and standard deviation 0.5
			\item Each agent has an  associated credit. This credit is used to get a credit based loan.
			\item Each agent has a parameter called quality, which tells us the quality of the crop that has been planted. Basically this parameter is a measure of how much produce the crop will give as compared to the optimal produce. So, if the agent was not able to hire labor, or the agent did not have water, or there was a famine then the quality decreases and hence the produce will also decrease. If the quality is 0, then the crop has essentially died and needs replacement. This parameter is between 0 and 1 and persists till the End-cycle of the crop has not been completed.
			\item Each agent has a parameter called income-expectation. This basically tells the agent what income can be expected from a particular crop. It is initialized based on the MSP/FRP/market price. This is gathered from the previous experience of the agent. It is set as the income gathered from the crop within the time period of 2 time steps from the harvest divided by the stock produced,
			\item Each agent has a parameter called upper-limit for land. This parameter basically tells the agent how much land is to be used to make it profitable to plant the crop. For a crop that has been planted, after the first harvest happens, we take the savings for some duration of time (in this case, the duration is equal to the harvest-cycle of the crop) prior to the current time and fit a line and record its slope. The upper-limit is updated as follows.
				\begin{align}
					upper\_limit[i]=min(land\_allocated[i]*(1+2*(\tan^{-1} m)/\pi),land)
				\end{align}
				In the above equation $m$ is the slope of the line that is fit and $i\in$ crops cultivated within the past 2 time steps. So, the basic idea is that if we have already planted crop $i$ in $land\_allocated[i]$ out of the available land, then if the saving dropped in the next few time steps(equal to the harvest-cycle of the crop), then the crop should be penalized as it was not sustainable. The penalty should be proportional to the drop of the savings. So, a line is fit for savings for the last few time steps(equal to the harvest-cycle of the crop) and then the angle is calculated, so that we can use the upper bound from the angle to determine the proportion. Similarly, if the savings rise, we need to reward the crop. $(1+2*(\tan^{-1} m)/\pi)$ does the part of rewarding and penalising for the crop. This is multiplied with the land allocated because we wish to limit the amount of land given to a crop that may not be rewarding, or increase the land given to a rewarding crop (upper bounded by the available land). This upper bound is used to see how much land we can allocate in future once the crop will be replaced which will be explained later.
			\item  Monthly expense (per-person-charge $*$ family size) is deducted at every monthly timestep from the total savings
			\item Whenever the land is no longer occupied, the agent calculates the water requirement for each crop based on the land planned to be used for planting the crop. Land planned to be used for planting the crop is estimated based on availabity of loan money, total savings, labor wages, labor requirement and estimated water price. Following is the equation used for  calculating the land to be used for each crop.
				\begin{align}
				\label{eq:landallt1}
					\begin{split}
						land\_allocated[i]=& max(min((ts+la-sb-ppc*fs*hc[i])/\\
								   &(ic[i]+ln[i]*lw[i]+wn[i]*wp+fpc[i]),ul[i]),0)
					\end{split}
				\end{align}
				Where $ts$ is total savings, $la$ is the loan amount estimated, $sb$ is safety buffer, $ppc$ is per person charge, $fs$ is the family size, $hc$ is the harvest cycle of the crop, $ic$ is initial cost, $ln[i]$ is the labor need for the i$^{th}$ crop, similarly $lw$ is for labor wages, $wn$ is water need, $wp$ is previous water price, $fpc$ is the fertilizer and pesticide cost, $ul$ is for upper limit of land. Basically, what we are trying to do here is find the amount of money the agent can put to use when the estimated family expenses are deducted and add the loan amount that can be taken, this way we can get the money that can be used for farming. Now all that we need to do is find the maximum amount of land that can be used for each crop subject to the money constraint and the $upper\_limit$ constraint. As all the expenses for the crop are taken as money per unit land (labor per unit land, fertilizer, pesticide per unit land and water per unit land), by equating costs and money for farming and rearranging, we can get the maximum amount of land that can be put to use for a crop subject to the money constraint. Now, we take the minimum of this result and the $upper\_limit$ to enforce the upper limit constraint. Finally, the maximum with respect to 0 is just a safety constraint to ensure that if the total savings are not enough, the land allocated should not go below 0.
				The water reqirement is estimated as the product of land allocated for that crop and water requirement of that crop. The list is then sorted according to the estimated profits. The estimated profit is calculated by the following formula:
				\begin{align}
					\label{eq:profit}
					\begin{split}
						profit[i]=&(c[i]/hc[i])*p[i]*l\_all[i]*ie[i]-(l\_all[i]*(ic[i]+\\&(wn*wp)*c[i]+ln[i]*c[i]*lw[i]))
				\end{split}
				\end{align}
				Where $c[i]$ is the end-cycle of the i$^{th}$ crop, similarly $p$ is for produce of crop, $hc$ is the harvest cycle of the crop, $ie$ is income expectation, {\em l\_all} is land allocated, $ic$ is initial cost, $ln[i]$ is the labor need for the i$^{th}$ crop, similarly $lw$ is for labor wages, $wn$ is water need, $wp$ is previous water price.\\
				The above details are communicated to the water agent or the Type3 farmer for water loan.
			\item When the agent recieves water from a water lender, if the water lender specifies a crop, then the agent plants that crop. The water lender would take a percentage of produce proportional to the water lent. If there is a crop which does not need any additional water other than rain water and is more profitable than the crop the water lender suggested after paying the percentage of produce, the water lender is notified. If the water lender does not specify the crop, then the choice is based on the most profitable given the water. This check is made because based on the amount of water recieved, the most profitable crop based on the estimated profits list calculated previously may no longer be the same. The profit is calculated the same way as in equation $\eqref{eq:profit}$, except $land\_allocated$ for each crop will be multiplied with the ratio between water recieved and water requested $land\_allocated[i]*(min(water\ recived,water\ requested[i])/water\ requested[i])$

			\item At every time step, the savings and expenses are checked, if the savings are not sufficient the farmer applies for a loan. As the loan is of two types, the loan based on credit is given higher preference over collateral based loan. Based on the expense calculated the money is split among the two types of loans. The total expense is calculated as follows:
				\begin{align}
				\label{eq:expense}
				\begin {split}
				expense=(curr\_time-end\_time)*(lw*ln+wn*wp+ppc*fs+fpc)+ic
				\end{split}
				\end{align}
			Where {\em curr\_time} is the current time and {\em end\_time} is the time before the current crop's end-cycle will complete, $ic$ will be 0 if the crop has already been planted, $ln$ is the labor need for the selected crop, similarly $lw$ is for labor wages, $wn$ is water need, $wp$ is previous water price, $fpc$ is the fertilizer and pesticide cost, $ppc$ is per person charge, $fs$ is the family size.
			This amount is then first applied for credit based loan, and the remaining amount is applied for colloateral based loan. Here land is taken as collateral.
		\item When the amount of money for planting of a crop can be covered by the savings and the loan, the planting is done.
		\item At every time step the water expense is deducted from the total savings. If the crop is recommended by the water lender, then the water expense is 0, as the payment is taken in the form of percentage of produce. If the savings cannot cover the water expense completely, then the quality of crop is decreased as follows
			\begin{align}
				quality=max(quality-(1-water\_flexibility*(water\_used/water\_needed),0)
			\end{align}
		\item labor expense is also handled in a similar way.
		\item Fertilizer and pesticide cost is cut at every time step. Unlike labor or water expense if the number of times the expense is not paid (due to lack of money) is greater than the prone to pest factor of the crop, the quality is set to 0. This is done because unlike labor or water, if there is a lack of fertilizers or pesticides, the crop may get infected with pests and that will cause the crop to go waste instead of loss of quality of yield. So, setting quality as 0 says that the crop is no longer usable. So, it is like giving some set number of chances and if that limit is crossed then penalty is .loss of crop. These dependencies can be modelled more realistically if actual data is available.
\item When the harvest cycle is reached, then the produce is received as the product of quality and the produce under optimum conditions.
		\item At every time step, the loan is also repaid. The installments are calculated the usual way. First, the collateral based loan is repaid, and then the credit based loan is repaid. If the number of defaults of collateral based loan is greater than a limit (currently 4) then the collateral is seized and  any excess amount is transfered to the farmer. If the loan can be closed with the savings and still have enough money to cover the costs, the loan is closed. Preference is given to collateral based loan followed by credit based loan.
		\item The farmer then gives the produce to the intermediate entity, which is either the mill, or the market agent.
		\item The farmer also sets an upper bound on the money for storage. This is 10\% of the savings. If the price is lower than the average price of the past 5 time steps, then the farmer applies for storage.
		\end{itemize}
	\item \textbf{Type2 Farmer}
		\begin{itemize}
			\item Most of the attributes and the functions are the same as Type1 farmer.
			\item There is no need of additional water from a water lender in this case. So all the actions and parameters that have water as an input will be modified.\cite{watermarket}
			\item  Land is drawn drawn from a gaussian distribution with mean 3 ha and standard deviation of 1. \cite{criteriafarmers} 
			\item Total savings are from a gaussian distribution with mean 30 lakh Rupees and standard deviation of 5,00,000.\cite{criteriafarmers} 
			\item The per person charge is 8000 rupees, with the same increment factor of 0.0001.
			\item Here because water is sufficient, the process of requesting water from a water lender is not required. So, whenever the land is no longer occupied, the farmer does {\em land\_allocated} using the following equation.
				\begin{align}
					\begin{split}
						land\_allocated[i]=& max(min((ts+la-sb-ppc*fs*hc[i])/\\
						 &(ic[i]+ln[i]*lw[i]+fpc[i]),ul[i]),0)
					\end{split}
				\end{align}
				The variable names are same as in equation $\eqref{eq:landallt1}$
				After land\_allocated is calculated, then the profit for each crop is calculated using the following equation
				\begin{align}
					\begin{split}
						profit[i]=&(c[i]/hc[i])*p[i]*l\_all[i]*ie[i]-\\&(l\_all[i]*(ic[i]+ln()[i]*c()[i]*lw[i]))
					\end{split}
				\end{align}
				The variable naming is same as in equation $\eqref{eq:profit}$
				Then based on the profit the crop is selected and planted just  as for a Type1 farmer.
			\item For applying loans, only the equation of expense changes, the rest remain the same. The credit rating for the Type2 farmer is higher than Type1 farmer (twice the credit). The equation for expense changes as follows.
				\begin{align}
					expense=(curr\_time-end\_time)*(lw*ln+ppc*fs+fpc)+ic
				\end{align}
                    The naming of variables is same as in equation $\eqref{eq:expense}$
			\item The water no longer affects the quality of the crop.
		\end{itemize}
	\item \textbf{Type3 Farmer}
		\begin{itemize}
			\item Most of the attributes and actions of the agent are the same as Type2 farmer.
			\item The farmer can also act as a water lender. The difference here as compared to the water agent is that the agent takes a portion of the produce for the water provided rather than charging any money.\cite{watermarket}
			\item A portion of produce from the water loanee is taken as payment.
			\item As there are some crops that require some minimum amount of produce to be made before the intermediate agents can come and collect the produce from the locality, the Type3 farmer acts as an agent influencing the choice of crop. If a type3 farmer gets some requests from the type1 farmers for water, then the Type3 farmer can suggest a crop from his preference. When the request for water is made, the type1 farmer also gives a list of crops that he can plant and the amount of land that the farmer is willing to use for each crop. So, when selecting the crop to be suggested, the Type3 farmer first sorts the list according to the estimated produce of the Type1 farmers for the crop he prefers the most. Then checks if there is an allocation possible such that the minimum produce for the preferred crop is possible. If yes, then the suggestions are finalized asking the Type1 farmers to produce that crop, if the minimum produce is not being met, then the next preferred crop is taken into consideration. We are assuming that if the Type1 farmer agrees to take the suggestion, then he will not break the agreement.
			\item There are two ways the water can be allocated, either by specifying the choice of crop or without specifying the choice. When specifying the choice of crop, following is the algorithm to allocate the water.
		\end{itemize}
		\begin{algorithm}[H]
					priority\_of\_crops=argsort(crops.price*crops.produce)\;
					\For{i in priority\_of\_crops}{
						sum\_of\_produce=0\;
						initial\_water=water\_available\;
						order=argsort(t1farmers\_applied.estimated\_produce(i))\;
						\For{j in order}{
							\If{$water\_available > 0$}{
								\eIf{t1farmers\_applied[j].water\_requirement[i]$\leq$ water\_available}{
								allocation(t1farmers\_applied[j])={i,t1farmers\_applied[j].water\_requirement[i]}\;
								water\_available-=t1farmers\_applied[j].water\_requirement[i]\;
								sum\_of\_produce+=t1farmers\_applied[j].estimated\_produce[i]\;
							}{
							temp=t1farmers\_applied[j].estimated\_produce[i]*water\_available\;
							sum\_of\_produce+=temp/t1farmers\_applied[j].water\_requirement[i]\;
							allocation(t1farmers\_applied[j])=\{i,water\_available\}\;
								water\_available=0\;
							}
							}
						}

						\eIf{sum\_of\_produce $\geq$ minimum\_produce[i]}{
							return the allocation\;
						}{
							reset\_all(allocation in this iteration)\;
						water\_available=initial\_water\;
						sum\_of\_produce=0\;
						}
					} 
					\caption{Assigning water and crops by the agent}
		\end{algorithm}
		In the case where the Type1 farmer accepts a different agent, then the Type3 farmers that have water remaining will again follow the same procedure for the the Type1 farmers that have not been given water, and the minimum produce constraint will be adjusted accordingly.

	\item \textbf{Water agent}
		\begin{itemize}
			\item This agent is basically an alternative to the Type3 farmers for lending water to the Type1 farmers. So, unlike the Type3 farmers, they do not require a part of the produce as the payment, but a water price is set. In our case, if there is a water agent, then the water agent is given a higher preference.
			\item The water agent does not care how the water is used, i.e. the agent does not care which crop is grown.
		\end{itemize}
	\item \textbf{Loan agent}
		\begin{itemize}
			\item The loan agent is responsible for loaning money.
			\item In the case of a collateral based loan, after 4 defaults from a farmer, the collateral is encashed and the excess amount, if any, is returned to the farmer.
			\item In the case of a credit based loan, every time the installments are paid on time, the credit rating is increased by 2, but every time an installment is defaulted, the credit rating is decreased by 8. Also there is a penalty of 20\% of the installment defaulted.
		\end{itemize}
	\item \textbf{Mill agent}
		\begin{itemize}
			\item The agent has parameter {\em yield\_of\_juice}, {\em yield\_of\_molasses}, {\em yield\_of\_sugar}, \\{\em yield\_of\_ethanol\_from\_molasses} and {\em yield\_of\_ethanol\_from\_juice} which are the parameters for quantity of sugarcane juice per unit sugarcane, quantity of molasses per unit sugarcane, quantity of ethanol produced from unit molasses and quantity of ethanol produced from unit sugarcane juice.
			\item There is a cost associated for transforming sugarcane to juice and molasses, molasses to ethanol and juice to ethanol.
			\item There is a minimum amount of money that is kept aside for maintainance and personal expenses.
			\item Credit based loan is taken from the Loan agents when the savings are lower than the estimated money needed. The money taken as loan is the maximum amount that can be taken from the loan agent based on the credit rating.
			\item The acquisition of sugarcane is based on the price of sugar, ethanol requirement, estimated sugar requirement and ethanol price. Following is the set of equations used to get the amount.
				\begin{align}
						ethanol\_produce&=sc*yj*e*yej+sc*ym*yem 
                \end{align}
                \begin{align}
                        \label{eq:only_sugar}
						sugar\_produce&=sc*yj*(1-e)*ys 
                \end{align}
				Constrained on
				\begin{align}
					ethanol\_produce&=ethanol\_requirement\\
					sugar\_produce&\geq estimated\_sugar\_requirement
				\end{align}
				Where $sc$ is sugarcane input, $yj$ is {\em yield\_of\_juice}, $e$ is percentage of sugarcane juice to ethanol (between 0 and 1), $yej$ is {\em yield\_of\_ethanol\_from\_juice}, $ym$ is {\em yield\_of\_molasses}, $yem$ is {\em yield\_of\_ethanol\_from\_molasses} and $ys$ is {\em yield\_of\_sugar}. The value of $e$ is set to 0 if the cost of production of alcohol from juice is greater than the price of ethanol. In this case the ethanol produce is entirely based on molasses and sugarcane is acquired entirely based on eq \eqref{eq:only_sugar}. \\
				When we rearrange terms, we get an inequality in terms of sugarcane input and constants, (treating estimated sugarcane as constant for that time step). We then get the lower bound of the sugarcane input, which we then use to acquire sugarcane from the farmers.
			\item The acquired sugarcane is then subject to the above equations to get the $e$ value and then the corresponding products are made.
			\item The sugar is then passed on to the market agent, while ethanol is passed to a government agency, which in our case is modelled as an environment variable.
		\end{itemize}

	\item \textbf{Storage agent}
		\begin{itemize}
			\item Given the storage object, the agent determines the price of the storage of each crop in a manner similar to the price of each crop. The demand of previous time step is analogous to total requests of storage from each agent for the crop. The {\em total\_stock} is analogous to the storage space remaining for each type of crop.\cite{coldstorage}
			\item The agent also cleans out the stock that has expired, or spoilt due to unavoidable circumstances.
		\end{itemize}

	\item \textbf{Market agent}
		\begin{itemize}
			\item The market agent takes the produce from the farmers and  storage and assigns the price according to the following equation.
				\begin{align}
				\begin{split}
					price[i]=&crop\_mult\_factor[i]*(max((2*previous\_sale[i][-4:]/total\_stock[i][-4:])\\&*[0.4,0.3,0.2,0.1],0.1))^2
					\end{split}
				\end{align}
				crop\_mult\_factor is a parameter that is based on the crop's produce. It can be thought of as how much the customers value the produce.  
			\item The product is sold at this price to the customers. The demand of the customers will depend on the price.
		\end{itemize}
	\item \textbf{Consumer agent}
		\begin{itemize}
		\item This is the agent that consumes the commodity and creates demand. The need for each commodity is within a specified limit irrespective of the price of the commodity, for example sugar consumption from 2015-2019 has been in the range of 24.5 million tons to 26 million tons per year\cite{consumption}
			\item The consumer agent uses the price and then increases or decreases the quantity demanded based on the price difference from the previous time step. There is a limit to how much the demand can vary and if the demand in the past 4 time steps is crossing the usual cumulative demand of 4 time steps, the demand is reduced by a factor of the ratio of excess and usual demand. The opposite is done in the case of demand being lower than usual demand. For example, if the cumulative sale of a commodity for the past 4 time time steps is 12000 units and the usual demand is 10000, then the excess demand is 2000. So, the ratio of excess demand to the usual demand is 0.2. So we will reduce the demand by 20\% because of the excessive sale in previous time steps. Now, let us say that the price in the previous time step was 150, and the price in the current time step is 250, then the demand for the current time step will be decreased by the ratio of the price difference and the current price. So the demand for the current time step will be $(1-0.4)*usual\_demand$. Now because of the excessive sale, this demand will be reduced by 20\%. So the final demand for this time step will be $(1-0.2)*(1-0.4)*usual\_demand$
		\end{itemize}

	\item \textbf{Import-Export agent}
		\begin{itemize}
			\item It has a parameter called {\em factor\_of\_export} and {\em factor\_of\_import} which are used while importing and exporting each consumable product respectively.
			\item It has a parameter called {\em maximum\_export} and {\em maximum\_import} which give upper bounds for export and import respectively.
			\item When the price is higher than the usual price, then the commodity is imported proportionate to the difference in price and the factor is the {\em factor\_of\_import}.
			\item Similar is the case for exports. Export is done when the price falls below the usual price.
		\end{itemize}
	\item \textbf{Policy agent}
		\begin{itemize}
			\item The agent reduces the taxes and increases the upper limit of import and factor of import if the number of imports made are more than two in the last 5 time steps. The agent also increases the export tax of the commodity. Similarly for the opposite case.
		\end{itemize}
\end{enumerate}

\subsection{Simulation}
The simulation has been carried out by creating 10000 farmer agents, with 70\% belonging to Type1, 20\% belonging to Type2 and rest as Type3. Each simulation was carried out for 50 time steps, and in some cases 500 time steps, to further analyze the results. We wanted to game-theoretically analyze the impact of a variable such as export price for each commodity, maximum export quantity permitted or price of water loaned by water agents on all the agents. So, we ran simulations in batches, where in each batch we selected one variable and ran simulations with different values of that variable and kept all the other variables at default values. Graphs of average savings of a type of farmer have been plotted against the variable that has been varied in the simulations for each batch. We then used the results of simulations of each batch to evaluate the performance of various policies.  
The github repository containing the code is https://github.com/satyandraguthula/simapi